# AL Pictoris and FR Piscis: two regular Blazhko RR Lyrae stars

Pierre de Ponthière [1]
*15 Rue Pré Mathy, Lesve – Profondeville 5170 – Belgium;*
*address email correspondence to pierredeponthiere@gmail.com*

Franz-Josef (Josch) Hambsch [1,2,3]
*12 Oude Bleken, Mol, 2400 - Belgium*

Kenneth Menzies [1]
*318A Potter Road, Framingham MA,01701 – USA*

Richard Sabo [1]
*2336 Trailcrest Drive, Bozeman MT, 59718 - USA*

1 American Association of Variable Star Observers (AAVSO)
2 Bundesdeutsche Arbeitsgemeinschaft für Veränderliche Sterne e.V. (BAV), Germany
3 Vereniging Voor Sterrenkunde (VVS), Belgium

## Abstract

The results presented are a continuation of observing campaigns conducted by a small group of amateur astronomers interested in the Blazhko effect of RR Lyrae stars. The goal of these observations is to confirm the RR Lyrae Blazhko effect and to detect any additional Blazhko modulation which cannot be identified from all sky survey data-mining. The Blazhko effect of the two observed stars is confirmed, but no additional Blazhko modulations have been detected.

The observation of the RR Lyrae star AL Pic during 169 nights was conducted from San Pedro de Atacama (Chile). From the observed light curve, 49 pulsation maxima have been measured. Fourier analyses of (O-C), magnitude at maximum light ($M_{max}$) and the complete light curve have provided a confirmation of published pulsation and Blazhko periods, 0.548622 and 34.07 days, respectively. The second multi-longitude observation campaign focused on the RR Lyrae star FR Psc was performed from Europe, United States and Chile. Fourier analyses of the light curve and of 59 measured brightness maxima have improved the accuracy of pulsation and Blazhko periods which are 0.45568 and 51.31 days, respectively. For both stars, no additional Blazhko modulations have been detected.

## 1. Introduction

The data-mining of automated sky surveys like All Sky Automated Survey - ASAS (Wils and Sódor 2005) and Northern Sky Variability Survey - NSVS (Wils *et al.* 2006) is frequently used to discover new RR Lyrae stars and to measure their pulsation periods and for some of them to detect and measure Blazhko modulation. Automated sky surveys with their low sampling frequencies (at best one sample per night) generate sparse datasets. As a result, spectral analysis of the datasets is not always fully reliable. Some Blazhko modulation periods are published as uncertain and multiple Blazhko modulations are not always detected. To overcome those shortcomings, more observations of identified RR Lyrae stars are required. Detailed study of long time-series observations allows the identification of individual brightness maxima of the light curve and other light curve details. The spectral analysis of the rich time-series is able to characterize the Blazhko modulation(s) in detail.

The results of observing campaigns presented herein are related to two RR Lyrae stars exhibiting the Blazhko effect, AL Pic and FR Psc, discovered respectively from ASAS and NSVS surveys (Wils and Sódor 2005 and Wils *et al.* 2006).

The designation of AL Pic appeared in the General Catalogue of Variable Stars with the 79[th] Name List of Variable Stars (Kazarovets *et al.* 2008), and previously this star was identified as GSC 8082-0469, NSV 1700 and ASAS J044131-5216.6. From the All Sky Automated Survey database, Wils and Sódor (2005) measured a pulsation period of 0.54861 day and also a Blazhko period of 34.0 days. The results presented herein are derived from data gathered during 169 nights between July 17, 2012 and February 1, 2013. A total of 17,416 magnitude measurements covering 5.8 Blazhko cycles were collected. All the observations were made by Franz-Josef Hambsch using remotely a 40 cm f/6.8 telescope located in San Pedro de Atacama (Chile).

The designation of FR Psc appeared in the General Catalogue of Variable Stars with the first part of 80[th] Name List of Variable Stars (Kazarovets *et al.* 2011). This star was previously identified as GSC 0607-0591 and NSV 9149730. From the Northern Sky Variability Survey data (Wozniak et al. 2004), Wils *et al.* (2006) have measured a pulsation period of 0.45570 days and an uncertain Blazhko period of 55 days. Our observations were made between August 18, 2012 and December 30, 2013. During the 126 observation nights, a total of 12,653 observations have been made by Franz-Josef Hambsch from Cloudcroft (NM) and from San Pedro de Atacama (Chile), by Richard Sabo from Bozeman (Montana), Kenneth Menzies from Framingham (MA) and Pierre de Ponthière from Lesve (Belgium). The numbers of observations for the different locations are 3597, 7272, 1106, 146 and 532, respectively.

For images of both stars, dark and flat field corrections were performed with MAXIMDL software (Diffraction Limited, 2004), and aperture photometry was performed using LESVEPHOTOMETRY (de Ponthière, 2010), a custom software which also evaluates the SNR and estimates magnitude errors. The photometric observations of both stars are available in the AAVSO International Database (AAVSO, 2014).

AL Pic observations were performed with only a V filter and are not transformed to the standard system. The comparison stars are given in Table 1, their coordinates and magnitudes in B and V bands were obtained from the AAVSO's Variable Star Database (VSD). The observations have been reduced with C1 as a magnitude reference and C2 as a check star.

The observations of FR Psc were also performed with a V filter and are not transformed to the standard system.
The comparison star coordinates and magnitudes in B and V bands were extracted from the AAVSO APASS survey and are given in Table 8. All observations, except from Cloudcroft, were reduced with C1 as magnitude reference and C2 as check star. Cloudcroft observations were reduced with C2 and C3 as reference and comparison stars. A correction of 0.021 magnitude has been applied to Cloudcroft observations. This correction has been calculated from the magnitude difference of C3 when measured with C1 and C2 as magnitude references.

## 2. Light curve maxima analyses

The times of maxima of the light curves have been evaluated with custom software (de Ponthière, 2010) fitting the light curve with a smoothing spline function (Reinsch, 1967).

### 2.1 AL Pic

A total of 49 maxima have been observed for AL Pic. Table 2 provides the list of the observed maxima and Figure 1 shows the (O-C) and $M_{max}$ (Magnitude at Maximum) values. A linear regression of all available (O-C) values has provided a pulsation period of 0.548549 d (1.822858 d$^{-1}$). The (O-C) values have been re-evaluated with this new pulsation period and the pulsation ephemeris origin has been set to the highest recorded brightness maximum: HJD 2 456 154.7560. The new derived pulsation elements are:

$$HJD_{Pulsation} = (2\ 456\ 154.7560 \pm 0.0080) + (0.548549 \pm 0.000044)\ E_{Pulsation}$$

The derived pulsation period is in good agreement with the value of 0.54861 published by Wils and Sódor (2005). The folded light curve on the newly determined pulsation period is shown in Figure 2.

To determine the Blazhko period, Fourier analyses and sine-wave fittings of the (O-C) values and $M_{max}$ (Magnitude at Maximum) values were performed with PERIOD04 (Lenz and Breger 2005). These analyses were limited to the three first harmonic components and are given in Table 3. The frequency uncertainties have been evaluated from the Monte Carlo simulation module of PERIOD04. The Blazhko periods obtained from (O-C) and $M_{max}$ are $34.23 \pm 0.12$ and $34.03 \pm 0.07$ days, respectively, which are in reasonable agreement. On this basis the best Blazhko ephemeris is

$$HJD_{Blazhko} = 2456154.75 + (34.03 \pm 0.07)\ E_{Blazhko}$$

The origin has been selected as the epoch of the highest recorded maximum. The (O-C) and $M_{max}$ curves folded with this Blazhko period are given in Figure 3. The variations of (O-C) and $M_{max}$ over the Blazhko cycles are respectively 0.097 day and 0.639 mag.

## 2.2 FR Psc

For FR Psc, the 59 observed maxima are listed in Table 9 and the graphs of (O-C) and $M_{max}$ values versus time (HJD) are given in Figure 6. A pulsation period of 0.4555800 d (2.1945223 $d^{-1}$) was derived from a linear regression of the (O-C) values. The (O-C) values have been re-evaluated with this new pulsation period and the pulsation ephemeris origin has been set to the highest recorded brightness maximum: HJD 2 456 631.5796. The new derived pulsation elements are:

$$HJD_{Pulsation} = (2\ 456\ 631.5796 \pm 0.0021) + (0.4556800 \pm 0.0000042)\ E_{Pulsation}$$

The calculated pulsation period is very close to the value of 0.45570 day published by Wils *et al*. (2006). The Blazhko amplitude and phase modulations are clearly visible in Figure 7 which presents the graph of the light curve folded with the pulsation period.

The Blazhko period has been measured with PERIOD04 in the same way as for AL Pic. The results are given in Table 10. The Blazhko periods, from (O-C) and $M_{max}$ are 51.32 ± 0.05 and 51.35 ± 0.09 days, respectively and agree within the errors. The Blazhko period of 55 days reported as uncertain by Wils *et al*. (2006) is slightly longer and not compatible with our improved results. The best Blazhko ephemeris is:

$$HJD_{Blazhko} = 2\ 456\ 631.58 + (51.32 \pm 0.05)\ E_{Blazhko}$$

The (O-C) and $M_{max}$ curves folded with the Blazhko period are given in Figure 8. The (O-C) and $M_{max}$ curves are anti-correlated and their shapes are very similar to the corresponding curves of AL Pic. The variations of (O-C) and $M_{max}$ over the Blazhko cycles are 0.042 day and 0.396 mag, respectively.

# 3. Frequency spectrum analyses of the light curve

For both AL Pic and FR Psc, the primary pulsations and Blazhko frequency have been derived from $M_{max}$ and (O-C) analyses. The Blazhko modulation frequencies appear clearly in the spectrum of the complete light curves. The method used for spectrum analysis has already been detailed in other observation reports (de Ponthiere *et al*. 2014), but the method is nevertheless briefly recalled here.

The spectrum of a signal modulated in amplitude and phase is characterized by a pattern of peaks called multiplets at the positions $kf_0 \pm nf_B$ with k and n being integer numbers corresponding respectively to the harmonic and multiplet orders. The frequencies, amplitudes and phases of the multiplets have been obtained with PERIOD04 by performing a succession of Fourier analyses, pre-whitenings and sine-wave fittings. Only the harmonic and multiplet components having a signal to noise ratio (SNR) greater than 4 have been retained as significant signals. Tables 5 and 12, for AL Pic and FR Psc, respectively, provide the complete lists of Fourier components with their amplitudes, phases and uncertainties. For both stars, besides the pulsation frequency $f_0$ and harmonics $nf_0$, one series of triplets $nf_0 \pm f_B$ based on the Blazhko frequency $f_B$ has been found. The Blazhko frequencies and corresponding periods are tabulated in Tables 4 and 11 with their uncertainties. These Blazhko periods are close to the values obtained with the $M_{max}$ analysis given in Tables 3 and 10.

During the sine-wave fitting, the fundamental frequencies $f_0$ and triplet $f_0 + f_B$ have been left unconstrained and the other frequencies have been entered as combinations of these two frequencies. The uncertainties of frequencies, amplitudes and phases have been estimated by Monte Carlo simulations. The amplitude and phase uncertainties have been multiplied by a factor of two as it is known that the Monte Carlo simulations underestimate these uncertainties (Kolenberg *et al*. 2009). Tables 6 and 13 list for each harmonic the amplitude ratios $A_i/A_1$ and the ratios usually used to characterize the Blazhko effect, that is, $A_i^+/A_1$ ; $A_i^-/A_1$ ; $R_i = A_i^+ / A_i^-$ and asymmetries $Q_i = (A_i^+ - A_i^-) / (A_i^+ + A_i^-)$. In the present cases the side lobe amplitudes $A_i^-$ and $A_i^+$ are similar which leads to small values for the $Q_i$ asymmetry ratios. Except for some higher order triplets of AL Pic, the asymmetry ratios are positive ($A_i^+ > A_i^-$). For the majority of Blazhko stars, the $Q_i$ asymmetry ratios are larger and generally lie between 0.1 and 0.5 but smaller and negative values are not unusual (Alcock *et al.* 2003). The fact that first triplet values ($A_1^+, A_1^-$) are larger for AL Pic (0.39 and 0.36) than for FR Psc values (0.17 and 0.16) is due to the relative strength of Blazhko modulations and is consistent with the corresponding variations of the magnitudes at maximum light provided in the preceding section (0.639 and 0.396 mag).

## 4. Light curve variations over Blazhko cycle

Subdividing the data set into temporal subsets is a classical method to visualize and analyze the light curve variations over the Blazhko cycle. For both stars, ten temporal subsets corresponding to the different Blazhko phase intervals $\Psi_i$ (i = 0 , 9) have been created using the epochs of the highest recorded maxima as the origins of the first subset. The folded light curves for the ten subsets are presented in Figures 4 and 9. For AL Pic, the data cover around six successive Blazhko cycles and are relatively well distributed over the different temporal subsets and Blazhko cycles. The population percentage of data points varies between 7.7% and 12% for the ten temporal subsets and between 8.4% and 30% for the six Blazhko cycles. The first and second observation seasons of FR Psc cover around 2.6 and 2.3 Blazhko cycles, respectively. The observation data are also well distributed over the temporal subsets, the population varying between 6.4% and 16%.

From a visual inspection of Figures 4 and 9, it is clear that the light curves are only affected by a small scatter during the successive Blazhko cycles. This fact is not surprising, indeed for both stars only one Blazhko modulation frequency has been detected in the spectrum analysis. Fourier analyses and Least-Square fittings have been performed on the different temporal subsets. For the fundamental frequency and the first four harmonics the amplitudes $A_i$ and the epoch-independent phase differences ($\Phi_{k1} = \Phi_k - k\Phi_1$) are given in Tables 7 and 14 and plotted in Figures 5 and 10. The differences between maximal and minimal values of $A_1$ over the Blazhko cycle for AL Pic and FR Psc are 0.235 and 0.106 mag, respectively. The larger Ai value for AL Pic is a confirmation that the Blazhko modulation is stronger for AL Pic than for FR Psc.

## 5. Conclusions

The two analysis methods, maximum brightness and light curve Fourier analyses, have provided similar results for both stars. No multiple or irregular Blazhko modulations have been detected and for the two stars the light curves repeat from one cycle to another. The FR Psc Blazhko period was previously published as uncertain. The new measured period value of 51.31 ± 0.02 days removes the uncertainty. The period of 34.07 ± 0.02 days for AL Pic is in agreement with the previously published value. The objective of this small group of amateur astronomers is to observe and to analyze Blazhko RR Lyrae stars with the hope of finding stars affected by irregular or multiple Blazhko modulations. These two coordinated multi-longitude campaigns have not revealed such multiple modulations. However, observers should continue their regular and coordinated multi-longitude observations to precisely characterize Blazhko modulations in other RR Lyrae stars.

## Acknowledgements

Dr A. Henden, Director of AAVSO and the AAVSO are acknowledged for the use of AAVSOnet telescopes at Cloudcroft (New Mexico, USA). The AAVSO Charts and Sequence Team is thanked for preparing the comparison star sequences. The authors thank the referee for the comments and encouragements. This work has made use of The International Variable Star Index (VSX) maintained by the AAVSO and the SIMBAD astronomical database (http://simbad.u-strasbg.fr).

**Table 1. AL Pic- Comparison stars**

| Identification | AAVSO AUID | R.A. (2000) h m s | Dec (2000) ° ' " | B | V | B-V | |
|---|---|---|---|---|---|---|---|
| GSC 8082-440 | 000-BKP-193 | 4:40:50.7 | -52:22:27.4 | 12.608 | 12.006 | 0.602 | C1 |
| GSC 8082-564 | 000-BKP-194 | 4:42:14.3 | -52:23:12.9 | 13.252 | 12.590 | 0.662 | C2 |

**Table 2. AL Pic - Measured Maxima**

| Maximum HJD | Error | O-C (day) | E | Magnitude (V) | Error |
|---|---|---|---|---|---|
| 2456144.8755 | 0.0060 | -0.0059 | -18 | 13.2840 | 0.015 |
| 2456149.8190 | 0.0027 | 0.0003 | -9 | 12.9820 | 0.012 |
| 2456150.9137 | 0.0020 | -0.0022 | -7 | 12.9020 | 0.010 |
| 2456154.7560 | 0.0030 | 0.0000 | 0 | 12.8060 | 0.011 |
| 2456155.8556 | 0.0026 | 0.0024 | 2 | 12.8710 | 0.011 |
| 2456160.7891 | 0.0040 | -0.0014 | 11 | 13.1290 | 0.012 |
| 2456161.8840 | 0.0048 | -0.0037 | 13 | 13.1720 | 0.013 |
| 2456167.8557 | 0.0049 | -0.0664 | 24 | 13.3320 | 0.014 |
| 2456172.7790 | 0.0085 | -0.0804 | 33 | 13.3760 | 0.019 |
| 2456177.7657 | 0.0059 | -0.0310 | 42 | 13.3250 | 0.015 |
| 2456182.7317 | 0.0030 | -0.0023 | 51 | 13.0730 | 0.012 |
| 2456183.8278 | 0.0043 | -0.0034 | 53 | 13.0180 | 0.026 |
| 2456188.7707 | 0.0035 | 0.0022 | 62 | 12.8800 | 0.014 |
| 2456189.8658 | 0.0020 | 0.0001 | 64 | 12.8810 | 0.013 |
| 2456221.6849 | 0.0028 | 0.0010 | 122 | 12.8660 | 0.015 |
| 2456222.7875 | 0.0022 | 0.0065 | 124 | 12.8970 | 0.013 |
| 2456226.6283 | 0.0038 | 0.0071 | 131 | 13.0180 | 0.015 |
| 2456233.7125 | 0.0038 | -0.0403 | 144 | 13.2920 | 0.016 |
| 2456234.8038 | 0.0068 | -0.0462 | 146 | 13.3650 | 0.021 |
| 2456239.7063 | 0.0053 | -0.0810 | 155 | 13.3470 | 0.017 |
| 2456240.8087 | 0.0079 | -0.0758 | 157 | 13.4140 | 0.018 |
| 2456244.7041 | 0.0071 | -0.0205 | 164 | 13.4160 | 0.017 |
| 2456245.8124 | 0.0048 | -0.0094 | 166 | 13.3930 | 0.014 |
| 2456249.6738 | 0.0057 | 0.0119 | 173 | 13.1860 | 0.025 |
| 2456250.7666 | 0.0029 | 0.0075 | 175 | 13.1350 | 0.015 |
| 2456254.6044 | 0.0028 | 0.0052 | 182 | 12.8370 | 0.014 |
| 2456255.7034 | 0.0029 | 0.0070 | 184 | 12.8530 | 0.017 |
| 2456256.8042 | 0.0023 | 0.0106 | 186 | 12.8680 | 0.02 |
| 2456259.5492 | 0.0034 | 0.0127 | 191 | 12.9610 | 0.018 |
| 2456261.7435 | 0.0029 | 0.0126 | 195 | 13.0910 | 0.018 |
| 2456265.5621 | 0.0043 | -0.0089 | 202 | 13.2590 | 0.014 |
| 2456266.6495 | 0.0055 | -0.0187 | 204 | 13.2920 | 0.02 |
| 2456272.6187 | 0.0052 | -0.0839 | 215 | 13.3530 | 0.015 |
| 2456273.7370 | 0.0085 | -0.0628 | 217 | 13.4330 | 0.021 |
| 2456277.6083 | 0.0064 | -0.0316 | 224 | 13.4250 | 0.013 |
| 2456278.7047 | 0.0043 | -0.0324 | 226 | 13.4300 | 0.009 |
| 2456283.6765 | 0.0031 | 0.0021 | 235 | 13.1350 | 0.016 |
| 2456288.6169 | 0.0024 | 0.0052 | 244 | 12.8560 | 0.007 |
| 2456289.7104 | 0.0019 | 0.0015 | 246 | 12.8240 | 0.015 |
| 2456293.5592 | 0.0018 | 0.0102 | 253 | 12.9820 | 0.006 |
| 2456294.6572 | 0.0026 | 0.0110 | 255 | 13.0200 | 0.007 |
| 2456295.7542 | 0.0068 | 0.0108 | 257 | 13.1420 | 0.033 |
| 2456299.5782 | 0.0050 | -0.0053 | 264 | 13.2820 | 0.016 |
| 2456300.6648 | 0.0034 | -0.0159 | 266 | 13.2850 | 0.008 |
| 2456301.7665 | 0.0112 | -0.0114 | 268 | 13.3850 | 0.021 |
| 2456305.5711 | 0.0153 | -0.0469 | 275 | 13.4250 | 0.025 |
| 2456307.7435 | 0.0092 | -0.0688 | 279 | 13.4450 | 0.015 |
| 2456317.6883 | 0.0043 | 0.0014 | 297 | 13.1820 | 0.013 |
| 2456321.5373 | 0.0040 | 0.0102 | 304 | 12.9370 | 0.021 |

**Table 3. AL Pic Blazhko spectral components from light curve maxima**

| From | Frequency (cycle/days) | σ(d$^{-1}$) | Period (days) | σ(d) | Amplitude | Φ (cycle) | SNR |
|---|---|---|---|---|---|---|---|
| (O-C) values | 0.02922 | 10 10$^{-5}$ | 34.23 | 0.12 | 0.037 day | 0.758 | 21.1 |
| M$_{max}$ values | 0.02939 | 6 10$^{-5}$ | 34.025 | 0.07 | 0.270 mag | 0.207 | 36.9 |

**Table 4 AL Pic Triplet component frequencies and periods**

| Component | Derived from | Frequency (d$^{-1}$) | σ(d$^{-1}$) | Period (d) | σ(d) |
|---|---|---|---|---|---|
| f$_0$ | | 1.822749 | 4.5x10$^{-6}$ | 0.548622 | 1.4x10$^{-6}$ |
| f$_B$ | f$_0$ + f$_B$ | 0.029353 | 13x10$^{-6}$ | 34.07 | 0.02 |

**Table 5 AL Pic Multi-frequency fit results**

| Component | f(d$^{-1}$) | σ(f) | A$_i$ (mag) | σ(A$_i$) | Φ$_i$ (cycle) | σ(Φ$_i$) | SNR |
|---|---|---|---|---|---|---|---|
| fo | 1.822749 | 4.5x10-6 | 0.2790 | 0.0007 | 0.5947 | 0.0005 | 99.8 |
| 2fo | 3.645498 | | 0.0972 | 0.0008 | 0.5489 | 0.0013 | 38.5 |
| 3fo | 5.468248 | | 0.0512 | 0.0007 | 0.4973 | 0.0022 | 22.3 |
| 4fo | 7.290997 | | 0.0321 | 0.0008 | 0.4488 | 0.0033 | 14.5 |
| 5fo | 9.113746 | | 0.0197 | 0.0007 | 0.4206 | 0.0065 | 11.2 |
| 6fo | 10.936495 | | 0.0117 | 0.0008 | 0.3931 | 0.0102 | 7.4 |
| 7fo | 12.759244 | | 0.0075 | 0.0007 | 0.3313 | 0.0167 | 5.3 |
| 8fo | 14.581993 | | 0.0051 | 0.0007 | 0.2765 | 0.0215 | 4.3 |
| f0 + fb | 1.852103 | 13x10-6 | 0.1090 | 0.0007 | 0.9100 | 0.0012 | 39.1 |
| f0 - fb | 1.793396 | | 0.1006 | 0.0007 | 0.9945 | 0.0012 | 35.9 |
| 2fo + fb | 3.674852 | | 0.0797 | 0.0007 | 0.9162 | 0.0015 | 31.5 |
| 2fo - fb | 3.616145 | | 0.0703 | 0.0007 | 0.9495 | 0.0017 | 27.9 |
| 3fo + fb | 5.497601 | | 0.0504 | 0.0007 | 0.9529 | 0.0023 | 22.0 |
| 3fo - fb | 5.438894 | | 0.0460 | 0.0008 | 0.9602 | 0.0027 | 20.0 |
| 4fo + fb | 7.320350 | | 0.0271 | 0.0007 | 0.9422 | 0.0044 | 12.3 |
| 4fo - fb | 7.261643 | | 0.0233 | 0.0007 | 0.9605 | 0.0053 | 10.5 |
| 5fo + fb | 9.143099 | | 0.0129 | 0.0007 | 0.8866 | 0.0100 | 7.3 |
| 5fo - fb | 9.084393 | | 0.0142 | 0.0008 | 0.8988 | 0.0083 | 8.0 |
| 6fo + fb | 10.965848 | | 0.0102 | 0.0008 | 0.8041 | 0.0125 | 6.5 |
| 6fo - fb | 10.907142 | | 0.0107 | 0.0008 | 0.8459 | 0.0125 | 6.8 |
| 7fo + fb | 12.788598 | | 0.0073 | 0.0007 | 0.7938 | 0.0157 | 5.1 |

**Table 6. AL Pic Harmonic, Triplet amplitudes, ratios and asymmetry parameters**

| $i$ | $A_i/A_1$ | $A_i^+/A_1$ | $A_i^-/A_1$ | $R_i$ | $Q_i$ |
|---|---|---|---|---|---|
| 1 | 1.00 | 0.39 | 0.36 | 1.08 | 0.04 |
| 2 | 0.35 | 0.29 | 0.25 | 1.13 | 0.06 |
| 3 | 0.18 | 0.18 | 0.16 | 1.10 | 0.05 |
| 4 | 0.12 | 0.10 | 0.08 | 1.16 | 0.08 |
| 5 | 0.07 | 0.05 | 0.05 | 0.91 | -0.05 |
| 6 | 0.04 | 0.04 | 0.04 | 0.95 | -0.03 |
| 7 | 0.03 | 0.03 | - | - | - |
| 8 | 0.02 | - | - | - | - |

**Table 7. AL Pic Fourier coefficients over Blazhko cycle**

| $\Psi$ (cycle) | $A_1$ (mag) | $A_2$ (mag) | $A_3$ (mag) | $A_4$ (mag) | $\Phi_1$ (rad) | $\Phi_{21}$ (rad) | $\Phi_{31}$ (rad) | $\Phi_{41}$ (rad) |
|---|---|---|---|---|---|---|---|---|
| 0.0 - 0.1 | 0.444 | 0.212 | 0.134 | 0.071 | 3.297 | 2.501 | 5.225 | 1.596 |
| 0.1 - 0.2 | 0.376 | 0.167 | 0.086 | 0.049 | 3.593 | 2.543 | 5.340 | 1.978 |
| 0.2 -0.3 | 0.286 | 0.120 | 0.063 | 0.036 | 3.504 | 2.522 | 5.232 | 1.625 |
| 0.3 - 0.4 | 0.235 | 0.095 | 0.034 | 0.021 | 4.462 | 2.798 | 5.394 | 2.124 |
| 0.4 - 0.5 | 0.209 | 0.096 | 0.041 | 0.021 | 4.446 | 2.433 | 5.305 | 1.913 |
| 0.5 - 0.6 | 0.235 | 0.078 | 0.032 | 0.023 | 4.554 | 2.482 | 5.319 | 1.765 |
| 0.6 - 0.7 | 0.205 | 0.083 | 0.035 | 0.015 | 3.459 | 2.204 | 4.660 | 1.182 |
| 0.7 - 0.8 | 0.285 | 0.125 | 0.072 | 0.028 | 3.877 | 2.177 | 4.841 | 1.023 |
| 0.8 - 0.9 | 0.362 | 0.192 | 0.116 | 0.080 | 3.422 | 2.508 | 5.310 | 1.685 |
| 0.9 - 1.0 | 0.421 | 0.223 | 0.156 | 0.099 | 3.350 | 2.461 | 5.217 | 1.623 |

**Table 8. FR Psc- Comparison stars**
*(AAVSO Chart 8256CED)*

| Identification | AAVSO AUID | R.A. (2000) h m s | Dec (2000) ° ' " | B | V | B-V | |
|---|---|---|---|---|---|---|---|
| GSC 607-409 | 000-BKP-155 | 0:47:35.6 | 11:47:09.1 | 12.347 | 11.657 | 0.690 | C1 |
| GSC 607-679 | 000-BKP-158 | 0:47:54.0 | 11:42:16.4 | 14.017 | 13.310 | 0.707 | C2 |
| GSC 607-799 | 000-BKP-159 | | | 14.837 | 14.237 | 0.600 | C3 |

**Table 9. FR Psc - Measured Maxima**

| Maximum HJD | Error | O-C (day) | E | Magnitude (V) | Error | Location |
|---|---|---|---|---|---|---|
| 2456158.5795 | 0.0020 | -0.0042 | -1038 | 11.639 | 0.010 | 1 |
| 2456159.4882 | 0.0045 | -0.0069 | -1036 | 11.629 | 0.013 | 1 |
| 2456192.7269 | 0.0051 | -0.0328 | -963 | 11.691 | 0.009 | 4 |
| 2456195.9113 | 0.0048 | -0.0382 | -956 | 11.682 | 0.045 | 4 |
| 2456202.7678 | 0.0055 | -0.0169 | -941 | 11.738 | 0.017 | 4 |
| 2456203.6806 | 0.0026 | -0.0154 | -939 | 11.708 | 0.016 | 4 |
| 2456205.9714 | 0.0049 | -0.0030 | -934 | 11.703 | 0.017 | 4 |
| 2456213.7200 | 0.0024 | -0.0010 | -917 | 11.477 | 0.014 | 4 |
| 2456214.6314 | 0.0014 | -0.0010 | -915 | 11.414 | 0.011 | 4 |
| 2456218.7341 | 0.0028 | 0.0006 | -906 | 11.360 | 0.036 | 4 |
| 2456219.6433 | 0.0019 | -0.0015 | -904 | 11.343 | 0.010 | 4 |
| 2456259.7407 | 0.0031 | -0.0040 | -816 | 11.635 | 0.026 | 4 |
| 2456261.5653 | 0.0023 | -0.0021 | -812 | 11.564 | 0.014 | 4 |
| 2456264.7575 | 0.0024 | 0.00034 | -805 | 11.493 | 0.017 | 4 |
| 2456265.6685 | 0.0019 | -0.00002 | -803 | 11.456 | 0.013 | 4 |
| 2456270.6801 | 0.0018 | -0.00090 | -792 | 11.357 | 0.011 | 4 |
| 2456540.4398 | 0.0014 | -0.00379 | -200 | 11.503 | 0.008 | 1 |
| 2456546.8092 | 0.0013 | -0.01391 | -186 | 11.659 | 0.005 | 2 |
| 2456546.8095 | 0.0016 | -0.01361 | -186 | 11.685 | 0.009 | 5 |
| 2456547.7192 | 0.0023 | -0.01527 | -184 | 11.678 | 0.006 | 2 |
| 2456551.8039 | 0.0040 | -0.03169 | -175 | 11.721 | 0.014 | 2 |
| 2456556.8141 | 0.0040 | -0.03397 | -164 | 11.724 | 0.017 | 5 |
| 2456560.4719 | 0.0037 | -0.02161 | -156 | 11.713 | 0.009 | 1 |
| 2456566.8696 | 0.0021 | -0.00343 | -142 | 11.668 | 0.010 | 5 |
| 2456567.7806 | 0.0035 | -0.00379 | -140 | 11.636 | 0.005 | 2 |
| 2456568.6969 | 0.0025 | 0.00115 | -138 | 11.607 | 0.006 | 2 |
| 2456571.8873 | 0.0003 | 0.00179 | -131 | 11.538 | 0.004 | 5 |
| 2456572.8007 | 0.0013 | 0.00383 | -129 | 11.484 | 0.004 | 2 |
| 2456573.7104 | 0.0014 | 0.00217 | -127 | 11.471 | 0.005 | 2 |
| 2456574.6216 | 0.0018 | 0.00201 | -125 | 11.418 | 0.005 | 2 |
| 2456578.7206 | 0.0010 | -0.00011 | -116 | 11.374 | 0.005 | 2 |
| 2456588.7447 | 0.0011 | -0.00098 | -94 | 11.441 | 0.005 | 2 |
| 2456589.6550 | 0.0013 | -0.00204 | -92 | 11.480 | 0.005 | 2 |
| 2456589.6557 | 0.0007 | -0.00134 | -92 | 11.507 | 0.005 | 5 |
| 2456592.8414 | 0.0012 | -0.00540 | -85 | 11.584 | 0.005 | 5 |
| 2456593.7524 | 0.0015 | -0.00576 | -83 | 11.574 | 0.006 | 2 |
| 2456594.6623 | 0.0013 | -0.00722 | -81 | 11.623 | 0.006 | 3 |
| 2456594.6635 | 0.0022 | -0.00602 | -81 | 11.587 | 0.008 | 2 |
| 2456595.5729 | 0.0021 | -0.00798 | -79 | 11.636 | 0.003 | 2 |
| 2456599.6659 | 0.0016 | -0.01610 | -70 | 11.685 | 0.004 | 2 |
| 2456600.5695 | 0.0019 | -0.02386 | -68 | 11.694 | 0.004 | 2 |
| 2456604.6661 | 0.0038 | -0.02838 | -59 | 11.703 | 0.005 | 2 |
| 2456605.5741 | 0.0025 | -0.03174 | -57 | 11.704 | 0.005 | 2 |
| 2456607.3937 | 0.0035 | -0.03486 | -53 | 11.711 | 0.007 | 1 |
| 2456609.6729 | 0.0030 | -0.03406 | -48 | 11.713 | 0.008 | 2 |
| 2456614.7092 | 0.0021 | -0.01024 | -37 | 11.708 | 0.010 | 2 |
| 2456615.6255 | 0.0030 | -0.00530 | -35 | 11.709 | 0.010 | 2 |
| 2456616.5376 | 0.0024 | -0.00456 | -33 | 11.681 | 0.010 | 2 |
| 2456619.7317 | 0.0008 | -0.00022 | -26 | 11.614 | 0.005 | 5 |
| 2456620.6441 | 0.0019 | 0.00082 | -24 | 11.575 | 0.009 | 2 |
| 2456621.5584 | 0.0029 | 0.00376 | -22 | 11.556 | 0.013 | 2 |
| 2456625.6577 | 0.0012 | 0.00194 | -13 | 11.414 | 0.006 | 2 |
| 2456626.5695 | 0.0017 | 0.00238 | -11 | 11.420 | 0.009 | 2 |
| 2456631.5796 | 0.0012 | 0.00000 | 0 | 11.342 | 0.015 | 2 |
| 2456636.5929 | 0.0016 | 0.00082 | 11 | 11.373 | 0.011 | 2 |
| 2456641.6030 | 0.0014 | -0.00156 | 22 | 11.491 | 0.008 | 2 |
| 2456647.5211 | 0.0014 | -0.00730 | 35 | 11.628 | 0.006 | 2 |
| 2456657.5222 | 0.0027 | -0.03116 | 57 | 11.738 | 0.008 | 3 |

*Locations: 1 - Lesve (Belgium); 2 - Bozeman (MT); 3 – Framingham (MA);*
*4 - Cloudcroft (NM); 5 - San Pedro de Atacama (Chile)*

**Table 10. FR Psc Blazhko spectral components from light curve maxima**

| From | Frequency (cycle/days) | $\sigma(d^{-1})$ | Period (days) | $\sigma(d)$ | Amplitude | $\Phi$ (cycle) | SNR |
|---|---|---|---|---|---|---|---|
| (O-C) values | 0.01948 | $2 \times 10^{-5}$ | 51.324 | 0.053 | 0.015 day | 0.948 | 33.1 |
| $M_{max}$ values | 0.01947 | $3 \times 10^{-5}$ | 51.353 | 0.087 | 0.193 mag | 0.428 | 21.0 |

**Table 11. FR Psc Triplet component frequencies and periods**

| Component | Derived from | Frequency ($d^{-1}$) | $\sigma(d^{-1})$ | Period (d) | $\sigma(d)$ |
|---|---|---|---|---|---|
| $f_0$ | | 2.1945066 | $1.3 \times 10^{-6}$ | 0.4556833 | $2.7 \times 10^{-7}$ |
| $f_B$ | $f_0 + f_B$ | 0.019490 | $6.1 \times 10^{-6}$ | 51.31 | 0.016 |

**Table 12. FR Psc Multi-frequency fit results**

| Component | $f(d^{-1})$ | $\sigma(f)$ | $A_i$ (mag) | $\sigma(A_i)$ | $\Phi_i$ (cycle) | $\sigma(\Phi_i)$ | SNR |
|---|---|---|---|---|---|---|---|
| fo | 2.194507 | $6.7 \times 10^{-7}$ | 0.4261 | 0.0011 | 0.1956 | 0.0004 | 252.2 |
| 2fo | 4.389014 | | 0.1768 | 0.0011 | 0.7527 | 0.0011 | 139.5 |
| 3fo | 6.583521 | | 0.0901 | 0.0010 | 0.3300 | 0.0019 | 77.9 |
| 4fo | 8.778028 | | 0.0379 | 0.0010 | 0.9201 | 0.0039 | 34.8 |
| 5fo | 10.972536 | | 0.0121 | 0.0009 | 0.3977 | 0.0138 | 11.9 |
| 6fo | 13.167043 | | 0.0111 | 0.0008 | 0.8574 | 0.0147 | 13.1 |
| 7fo | 15.361550 | | 0.0115 | 0.0010 | 0.3822 | 0.0175 | 16.3 |
| 8fo | 17.556057 | | 0.0103 | 0.0010 | 0.9469 | 0.0177 | 17.0 |
| 9fo | 19.750564 | | 0.0087 | 0.0010 | 0.5409 | 0.0182 | 13.2 |
| f0 + fb | 2.213999 | $30 \times 10^{-7}$ | 0.0731 | 0.0011 | 0.7259 | 0.0019 | 43.7 |
| f0 - fb | 2.175015 | | 0.0664 | 0.0010 | 0.3003 | 0.0026 | 39.2 |
| 2fo + fb | 4.408506 | | 0.0627 | 0.0009 | 0.2775 | 0.0025 | 49.4 |
| 2fo - fb | 4.369522 | | 0.0591 | 0.0011 | 0.8392 | 0.0026 | 46.5 |
| 3fo + fb | 6.603013 | | 0.0608 | 0.0010 | 0.8647 | 0.0024 | 52.6 |
| 3fo - fb | 6.564030 | | 0.0543 | 0.0011 | 0.4355 | 0.0034 | 46.9 |
| 4fo + fb | 8.797520 | | 0.0391 | 0.0009 | 0.4688 | 0.0048 | 35.8 |
| 4fo - fb | 8.758537 | | 0.0357 | 0.0011 | 0.0446 | 0.0045 | 32.7 |
| 5fo + fb | 10.992027 | | 0.0264 | 0.0011 | 0.0512 | 0.0061 | 26.1 |
| 5fo - fb | 10.953044 | | 0.0216 | 0.0010 | 0.6616 | 0.0069 | 21.3 |

| | | | | | | |
|---|---|---|---|---|---|---|
| 6fo + fb | 13.186535 | 0.0161 | 0.0010 | 0.6491 | 0.0090 | 19.0 |
| 6fo - fb | 13.147551 | 0.0122 | 0.0010 | 0.2430 | 0.0158 | 14.4 |
| 7fo + fb | 15.381042 | 0.0087 | 0.0011 | 0.2217 | 0.0168 | 12.4 |
| 7fo - fb | 15.342058 | 0.0065 | 0.0010 | 0.8239 | 0.0263 | 9.2 |
| 8fo + fb | 17.575549 | 0.0040 | 0.0011 | 0.7335 | 0.0424 | 6.6 |
| 8fo - fb | 17.536565 | 0.0022 | 0.0011 | 0.3048 | 0.0793 | 3.7 |

**Table 13. FR Psc Harmonic, Triplet amplitudes, ratios and asymmetry parameters**

| $i$ | $A_i/A_1$ | $A_i^+/A_1$ | $A_i^-/A_1$ | $R_i$ | $Q_i$ |
|---|---|---|---|---|---|
| 1 | 1.00 | 0.17 | 0.16 | 1.10 | 0.05 |
| 2 | 0.41 | 0.15 | 0.14 | 1.06 | 0.03 |
| 3 | 0.21 | 0.14 | 0.13 | 1.12 | 0.06 |
| 4 | 0.09 | 0.09 | 0.08 | 1.09 | 0.05 |
| 5 | 0.03 | 0.06 | 0.05 | 1.23 | 0.10 |
| 6 | 0.03 | 0.04 | 0.03 | 1.32 | 0.14 |
| 7 | 0.03 | 0.02 | 0.02 | 1.34 | 0.14 |
| 8 | 0.02 | 0.01 | 0.01 | 1.80 | 0.29 |
| 9 | 0.02 | - | - | - | - |

**Table 14. FR Psc Fourier coefficients over Blazhko cycle**

| $\Psi$ (cycle) | $A_1$ (mag) | $A_2$ (mag) | $A_3$ (mag) | $A_4$ (mag) | $\Phi_1$ (rad) | $\Phi_{21}$ (rad) | $\Phi_{31}$ (rad) | $\Phi_{41}$ (rad) |
|---|---|---|---|---|---|---|---|---|
| 0.0 - 0.1 | 0.496 | 0.234 | 0.159 | 0.084 | 3.977 | 2.276 | 4.728 | 1.092 |
| 0.1 - 0.2 | 0.462 | 0.211 | 0.134 | 0.068 | 4.149 | 2.311 | 4.792 | 1.151 |
| 0.2 - 0.3 | 0.441 | 0.179 | 0.106 | 0.054 | 3.766 | 2.357 | 4.850 | 1.197 |
| 0.3 - 0.4 | 0.409 | 0.173 | 0.093 | 0.044 | 4.379 | 2.322 | 4.822 | 1.201 |
| 0.4 - 0.5 | 0.394 | 0.170 | 0.098 | 0.050 | 4.540 | 2.324 | 4.897 | 1.207 |
| 0.5 - 0.6 | 0.390 | 0.161 | 0.094 | 0.047 | 4.646 | 2.309 | 4.934 | 1.222 |
| 0.6 - 0.7 | 0.402 | 0.163 | 0.082 | 0.038 | 4.437 | 2.282 | 4.929 | 1.409 |
| 0.7 - 0.8 | 0.418 | 0.180 | 0.087 | 0.035 | 4.200 | 2.304 | 4.767 | 1.021 |
| 0.8 - 0.9 | 0.456 | 0.212 | 0.132 | 0.068 | 4.082 | 2.238 | 4.629 | 0.921 |
| 0.9 - 1.0 | 0.491 | 0.240 | 0.170 | 0.108 | 3.979 | 2.273 | 4.712 | 1.046 |

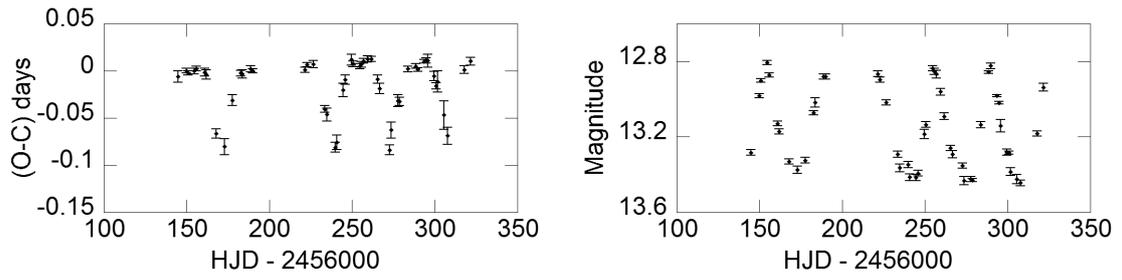

Figure 1. AL Pic O-C (days) and Magnitude at maximum

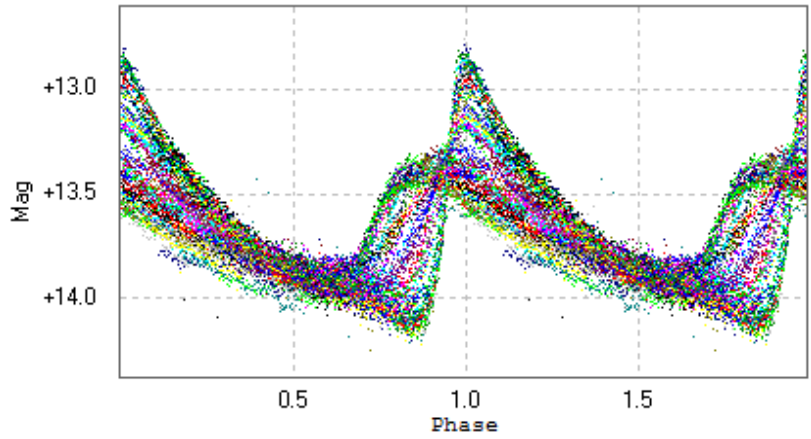

Figure 2. AL Pic light curve folded with pulsation period

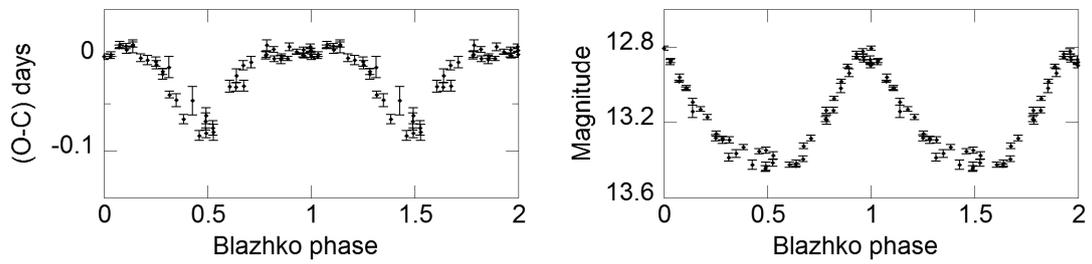

Figure 3. AL Pic O-C and Magnitude at maximum folded with Blazhko period

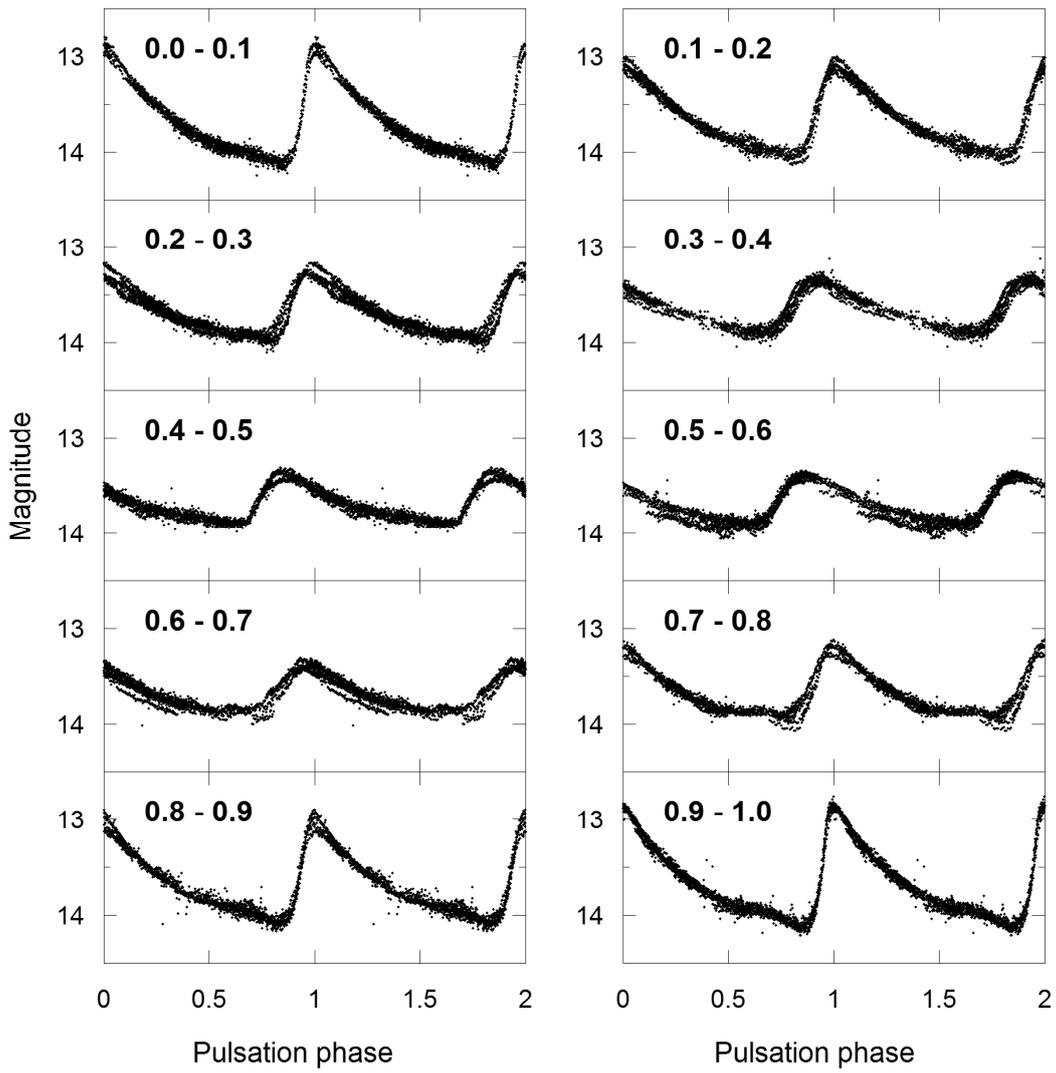

Figure 4. AL Pic light curves (magnitude vs. pulsation phase)
for the ten temporal subsets based on Blazhko period.

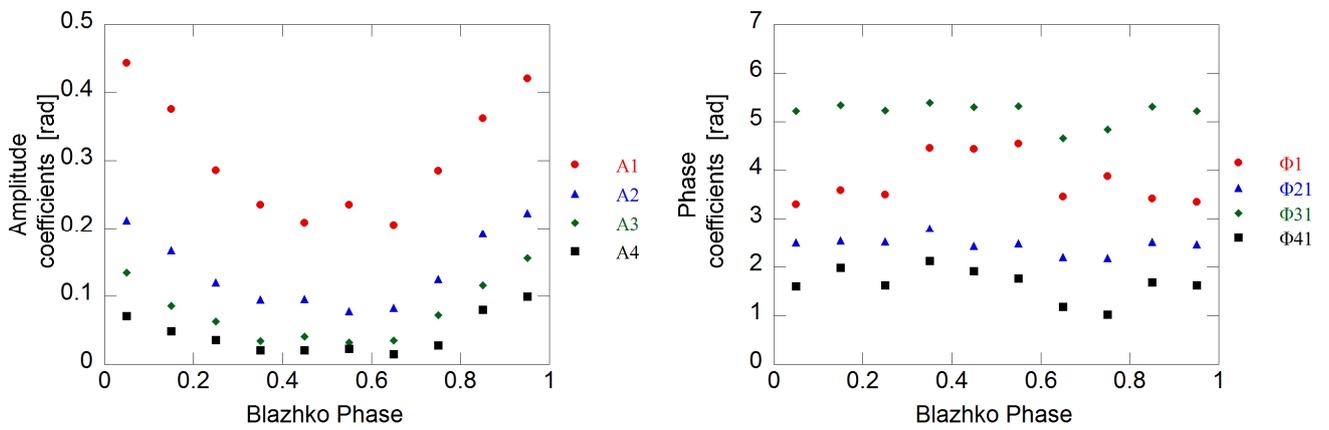

Figure 5. Al Pic Fourier amplitude and phase coefficients (mag)
for the ten temporal subsets based on a Blazhko period.

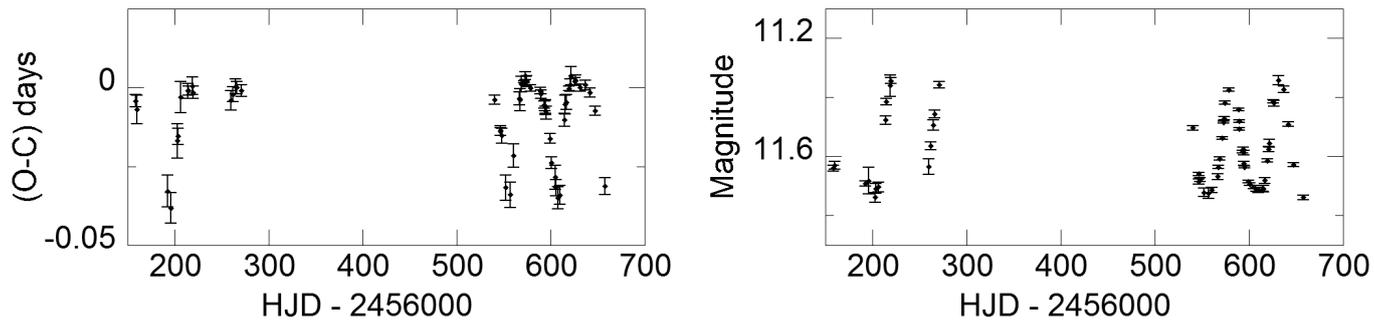

Figure 6. **FR Psc** O-C (days) and Magnitude at maximum

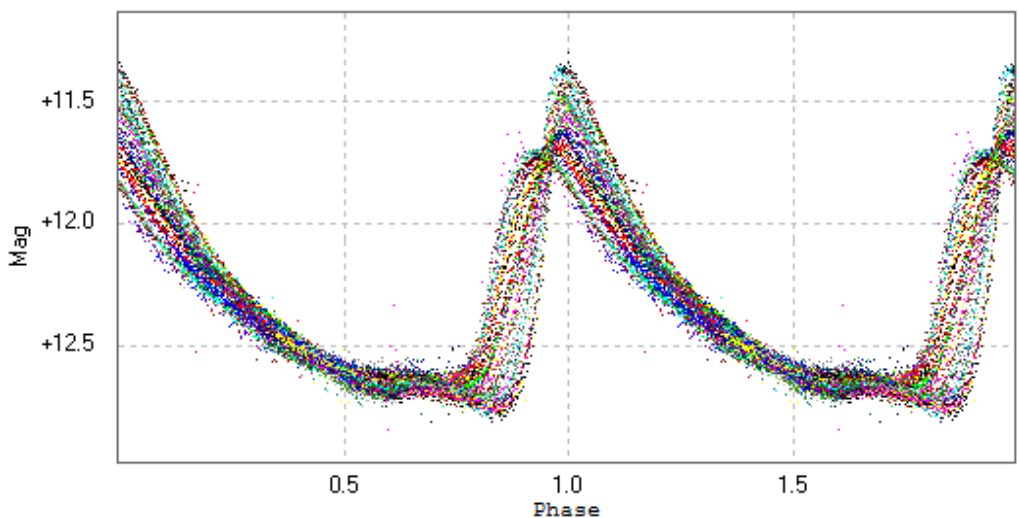

Figure 7. **FR Psc** light curve folded with pulsation period

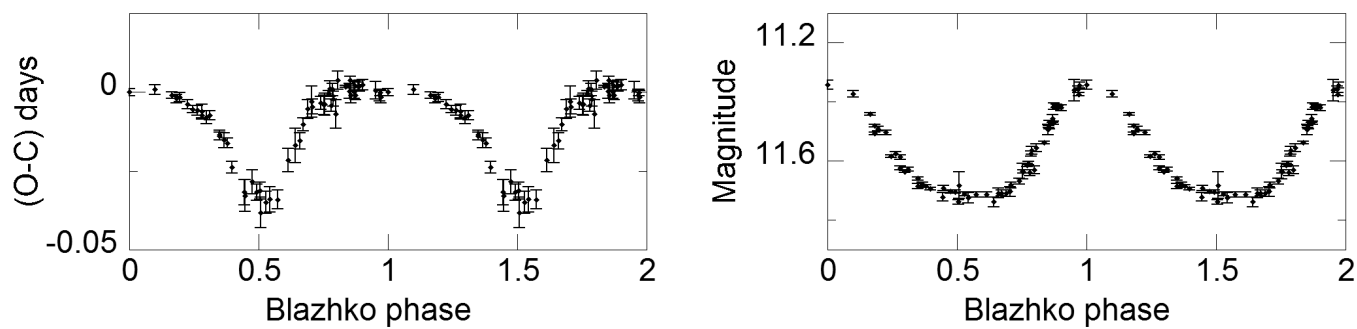

Figure 8. **FR Psc** O-C and Magnitude at maximum folded with Blazhko period

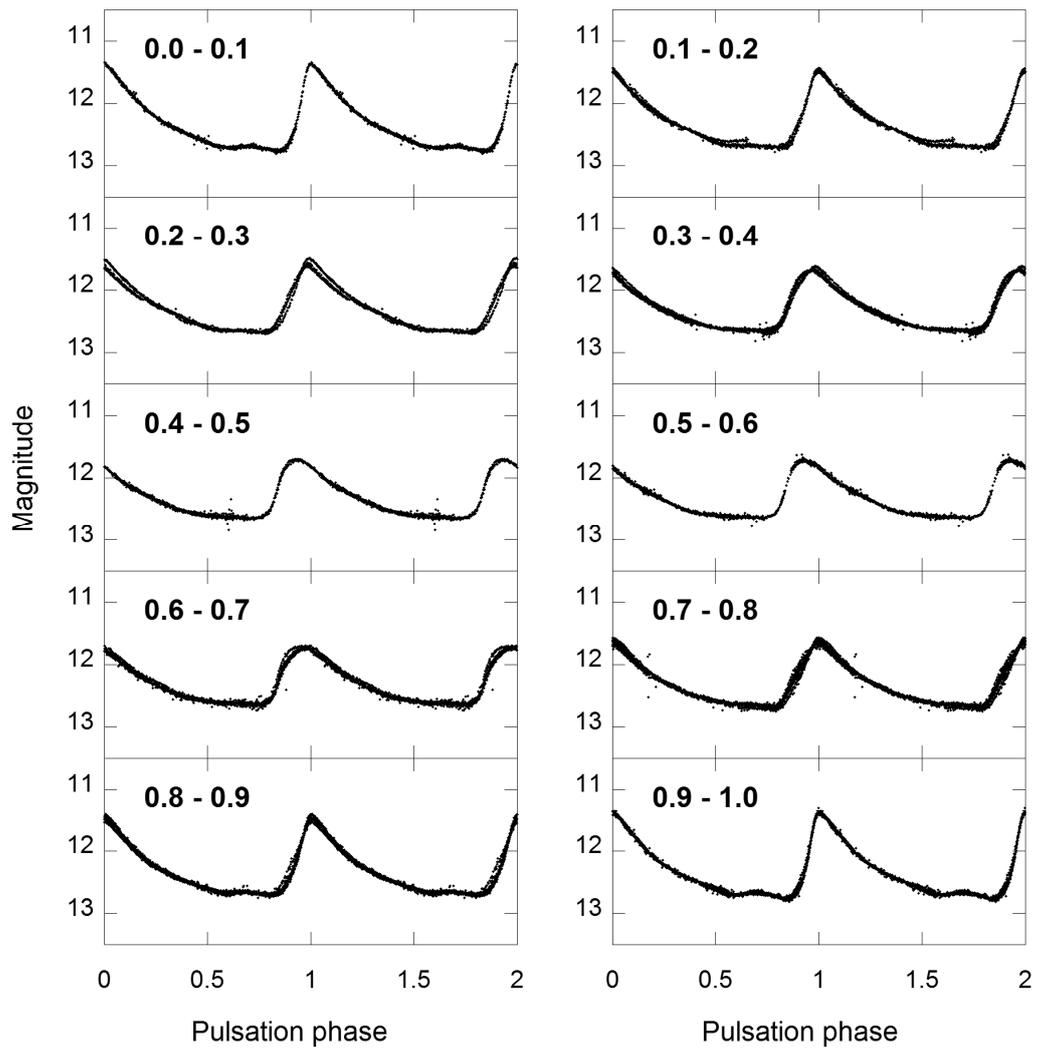

Figure 9. **FR Psc** light curves (magnitude vs. pulsation phase) for the ten temporal subsets based on Blazhko period.

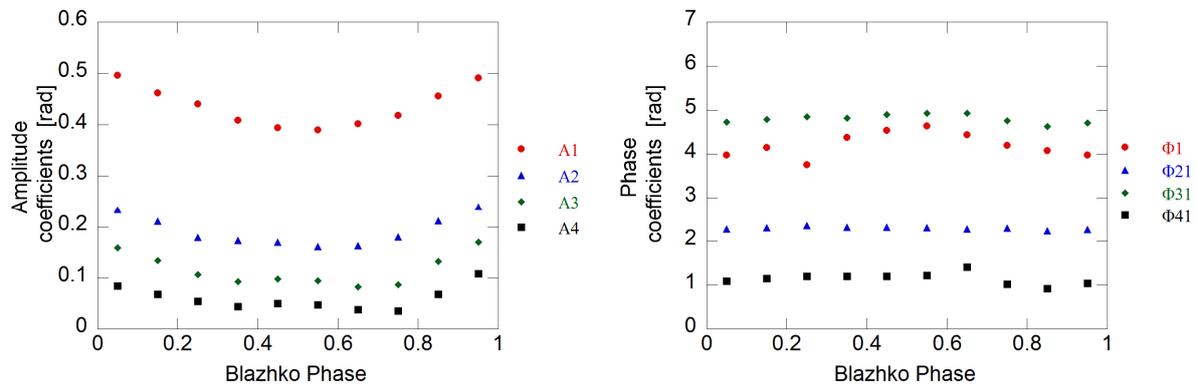

Figure 10. **FR Psc** Fourier amplitude and phase coefficients (mag) for the ten temporal subsets based on a Blazhko period.

14